\documentclass[%
 reprint,
 amsmath,amssymb,
 aps,
]{revtex4-1}

\usepackage{graphicx,subfigure }% Include figure files
\usepackage{dcolumn}% Align table columns on decimal point
\usepackage{bm}% bold math
\begin{document}

\preprint{APS/123-QED}

\title{Melting/freezing transition in polydisperse Lennard-Jones system: Remarkable agreement between predictions of 
inherent structure, bifurcation phase diagram, Hansen-Verlet rule and Lindemann criteria}

\author{Sarmistha Sarkar$^{1, 2}$, Rajib Biswas$^1$, Partha Pratim Ray$^2$ and Biman Bagchi$^{1,}$}
\email{Corresponding author: bbagchi@sscu.iisc.ernet.in}
\affiliation{$^1$Solid State and Structural Chemistry Unit, Indian Institute of Science, Bangalore - 560012}
\affiliation{$^2$Department of Physics, Jadavpur University, Kolkata - 700032, India}
\date{\today}
\begin{abstract}
We use polydispersity in size as a control parameter to explore certain aspects of melting and freezing transitions
in a system of Lennard-Jones spheres. Both analytical theory and computer simulations are employed to establish a 
potentially interesting relationship between observed terminal polydispersity in Lennard-Jones polydisperse spheres
and prediction of the same in the integral equation based theoretical analysis of liquid-solid transition. As we 
increase polydispersity, solid becomes inherently unstable because of the strain built up due to the size disparity.
This aspect is studied here by calculating the inherent structure (IS) calculation. With polydispersity at constant 
volume fraction we find initially a sharp rise of the average IS energy of the crystalline solid until transition 
polydispersity, followed by a cross over to a weaker dependence of IS energy on polydispersity in the amorphous state.
This cross over from FCC to amorphous state predicted by IS analysis agrees remarkably well with the solid-liquid 
phase diagram (with extension into the metastable phase) generated by non-linear integral equation theories of freezing.
Two other well-known criteria of freezing/melting transitions, the Hansen-Verlet rule of freezing and the Lindemann 
criterion of melting are both shown to be remarkably in good agreement with the above two estimates. Together they 
seem to indicate a small range of metastability in the liquid-solid transition in polydisperse solids.
\end{abstract}

\maketitle

\section{ Introduction}
In earlier studies of polydisperse system it has been established that homogeneous crystallization never takes place above an upper 
limit of the polydispersity, known as terminal polydispersity~\cite{sneha_prl,lacks_jcp,kofke_pre_99,megen_pre_01,frenkel_nat_01,
hansen_mcdonald,shouji_jmol}. Recently, an application of density functional theory (DFT) theory of freezing~\cite{tvr_prb_79,tvr_prl_82}
of hard sphere fluid by Chaudhuri et al. found a terminal polydispersity of 0.048, followed by a reentrant melting at larger density
~\cite{chandan_prl}. Several studies showed the existence of terminal polydispersity as well as reentrant melting~\cite{terada_jpcb_05,
bartlett_prl_99} on hard sphere polydisperse systems. Experimental studies, however, observed a nearly universal value of 0.12 for 
the terminal polydispersity~\cite{phan_jcp,megen_langmuir}. A number of theoretical and experimental techniques have been utilized 
to understand freezing/melting transition~\cite{haymet_oxtoby_1,haymet_oxtoby_2,oxtoby_haymet,feder_prl_70,schneider_pra,gibbs_jcp,
simon_prl_93,hoffman_pre,pgw_stoessel_prl,pgw_stoessel_jcp,langer_pre,tanaka_jcp_13,buff_jcp_76}. Although an analytical study of 
phase transition of polydisperse system is challenging, there have been a number of studies on fluid-fluid and the fluid-solid transition
~\cite{sollich_jcp_06}. Both the freezing and melting transition of polydisperse colloidal system have been examined by 
L\"{o}wen et al~\cite{simon_prl_93,hoffman_pre}. A series of molecular-dynamics calculations has been carried out to examine melting
in the classical Gaussian core model~\cite{stillinger_weber_prb}.

In the melting of solids when molecules/atoms interact with each other via a central potential, two empirical criteria have been 
used extensively to predict transitions. From the melting side one uses the Lindemann criterion~\cite{lindemann} which states that at melting 
temperature, the root mean square displacement exceeds 0.1 of the cell length. The second criterion is the Hansen-Verlet rule of 
crystallization~\cite{hansen_pr} that requires the value of the first peak of static structure factor $S(k)$ to exceed 2.85 for 
freezing. Both these simple criteria are found to hold good for different types of interactions and hence these criteria are 
considered universal. Lindemann criteria of melting can be investigated using lattice dynamical theory. On the other hand, justification 
of the Hansen-Verlet rule of crystallization comes from the density functional theory of freezing~\cite{tvr_prb_79,tvr_prl_82,haymet_oxtoby_1,
haymet_oxtoby_2,oxtoby_haymet,pgw_stoessel_prl}. These theories have also been used to understand metastability and nucleation in the 
liquid-solid phase transition of the polydisperse system.

In the present work we apply both theory and computer simulation to establish a relationship between integral equation theory of freezing 
and terminal polydispersity during the first order liquid-solid phase transition in polydisperse Lennard-Jones spheres. The inherent 
structure analysis~\cite{wales} further enlightens the understanding of the relationship between transition and stability of the solid.

Integral equation based theoretical analysis of freezing transition has been a subject of great interest for many years. The most 
successful theory of freezing, the Ramakrishnan-Yussouff theory~\cite{tvr_prl_82,tvr_prb_79} is also based, at the core, on integral 
equations that relate the inhomogeneous single particle density to two particle direct correlation function, \textit{c(r)}. This 
theory employs an expansion of density in a Fourier series where order parameters are the density evaluated at the reciprocal 
lattice vector components. The resulting equations are solved along with the thermodynamics conditions of equality of chemical 
potential and equality of pressure.

In fact, relationship between solutions of integral equation theories and liquid-solid transition has been investigated extensively. 
Many of the previous studies focused on the limiting liquid density beyond which liquid becomes unstable with respect to periodic 
density waves of the solid. In the analysis of Lovett~\cite{lovett_jcp_77}, this instability point was identified as 
$\rho _{l} \vec{C}_{2} \left(\vec{G}\right)=1$. Similar conclusion was reached by Munakata \textit{et al}~\cite{munakata_1,munakata_2}
in an interesting approach using the correct non-linear integral theories. Rice \textit{et al} built on these previous analyses, but
made the analysis robust by using proper order parameters. In the analysis of Rice \textit{et al}~\cite{bcr_1}, $\lambda _{\vec{G}} =1$is 
identified as the limit of instability or spinodal point, where $\lambda _{\vec{G}} =\rho _{S} \, C(G)$, with $\rho _{S} $as the 
density of the emerging solid phase.

In addition, enormous theoretical approaches are used to describe the freezing transition directly in terms of the bifurcation of 
solutions for single density from the homogeneous liquid branch to inhomogeneous branch. This approach also applies essentially 
the same integral equation that relates the inhomogeneous singlet density to the pair direct correlation function~\cite{lovett_jcp_77,
bcr_1,buff_jcp_80,xu_rice_pre_08,bcr_2,xu_rice_pre_11,bcr_3}. Interestingly, a terminal point in the bifurcation diagram is achieved 
where the density of the liquid phase is close to the dense random close packed state of the hard sphere liquid and the density of 
the solid phase is close to crystal closed packed values~\cite{bcr_1}. This point can be interpreted as the signature of the end 
of possible compression in the system.

In the elegant approach of the theory of freezing by Ramakrishnan and Yussouff~\cite{tvr_prl_82,tvr_prb_79}, thermodynamic potential
of the system is computed as a function of the order parameters proportionl to the periodic lattice components of the singlet density.
This theory employs a nearly exact expression for inhomogeneous density field derived from the density functional theory of statistical
mechanics. Free energy is expressed in terms of density fluctuations around the bulk liquid density with n-particle direct correlation
functions as the expansion coefficients, as elaborated below.

\section{Theoretical background}
In accordance with the density functional theory of freezing, the density of inhomogeneous solid can be expressed in terms of order 
parameters $\phi _{0} $ and $\left\{\phi _{\vec{G}} \right\}$ in the following fashion
\begin{equation} \label{eq1} 
\rho _{s} \left(\vec{R}\right)=\rho _{l} \left(1+\phi _{0} \right)+\rho _{l} \sum _{\vec{G}}\phi _{\vec{G}} \exp \left(i\vec{G}\cdot \vec{R}\right), 
\end{equation} 
where $\rho _{s} \left(\vec{R}\right)$ is the inhomogeneous solid density, $\rho _{l} $ is the liquid density and $\left\{\vec{G}\right\}$ is 
the set of reciprocal lattice vectors of the chosen lattice. The order parameters $\phi _{0} $ and $\left\{\phi _{\vec{G}} \right\}$ are 
the expansion coefficients of the inhomogeneous solid density, defined quantitatively below. These quantities can be quantified
in the following fashion,
\begin{equation} \label{eq2} 
\rho _{S} =\frac{1}{\Delta } \int _{\Delta }\rho _{S} \left(\vec{R}\right) d\vec{R}, 
\end{equation} 
where integration is over the lattice cell of volume $\Delta.$
Fractional density change $\phi _{0} $ in the transition can clearly be written as $\phi_{0} =\left(\rho_{s} -\rho_{l} \right)/\rho_{l}$. 
Structural order parameters $\phi _{\vec{G}} $ can be obtained by solving the following integral equations as,
\begin{equation} \label{eq4} 
\phi _{\vec{G}} =\frac{1}{\Delta } \int \frac{\Delta \rho \left(\vec{R}\right)}{\rho _{l} }  e^{-i\vec{G}\cdot \vec{R}} d\vec{R}. 
\end{equation} 
One can introduce a scaling transformation that brings out an interesting generality of the freezing theory as,
\begin{equation} \label{eq5} 
\lambda _{\vec{G}} =\rho _{s} \vec{C}_{2} \left(\vec{G}\right), 
\end{equation} 
\begin{equation} \label{eq6} 
\psi _{\vec{G}} =\frac{\rho _{l} }{\rho _{s} } \phi _{\vec{G}} . 
\end{equation} 
where $\vec{C}_{2} \left(\vec{G}\right)=\int dRC(R) e^{i\vec{G}\cdot \vec{R}} .$ 
Therefore, $\lambda _{\vec{G}} $ and $\psi _{\vec{G}} $ are both dimensionless quantities.
$\vec{C}_{2} \left(\vec{G}\right)$ is the Fourier transform of the pair direct correlation function, evaluated at $\vec{G}$.
The above transformations lead to the following expression $\psi _{\vec{G}_{n} } ,$
\begin{equation} \label{eq7} 
\psi _{\vec{G}_{n} } =\frac{\int _{\Delta }d\vec{R}_{1} \xi _{\vec{G}_{n} } \left(\vec{R}_{1} \right)
                      \exp \left[\sum _{\vec{G}}\psi _{\vec{G}} \lambda _{\vec{G}}  \xi _{\vec{G}} 
                      \left(\vec{R}_{1} \right)\right] }{\int _{\Delta }d\vec{R}_{1} \exp 
                      \left[\sum _{\vec{G}}\psi _{\vec{G}} \lambda _{\vec{G}}  \xi _{\vec{G}} \left(\vec{R}_{1} \right)\right]} . 
\end{equation} 
Here $\vec{G}_{n} $ is the n-th reciprocal lattice vector, $\xi _{\vec{G}_{n} } $is the periodic density wave corresponds to the set 
of $\left\{\vec{G}\right\}$ that has the same magnitude.
So,
\begin{equation} \label{eq8} 
\xi _{G_{\alpha } } (R)=\sum _{\left\{G_{\alpha } \right\}}e ^{iG_{\alpha } .\overrightarrow{R}} . 
\end{equation} 
Note that Eq.~\ref{eq7} has a quasi-universal character, as noted by Rice\textit{ et al}~\cite{bcr_1}. Equation~\ref{eq7}
does not require any information about direct correlation function or density of the liquid and the solid phases. Only the lattice type needs
to be specified. For the calculation of phase transition parameters, Eq.~\ref{eq7} needs to be supplied by the equality of 
thermodynamic potential as $\Omega _{s}-\Omega _{l} =0.$ 

Note that Eq.~\ref{eq7} can also be derived by combining equations of Ramakrishnan-Yussouff and of Haymet-Oxtoby 
(see Eq.~\ref{eq9} below). The scaling transformation allows one to solve for $\psi _{\vec{G}} $ as a function of $\lambda _{\vec{G}} $without 
requiring the input of direct correlation function, and the functional form has an interesting character as shown below.

Using the density functional theory, one can obtain the following expression of the grand canonical potential difference between solid and liquid as,
\begin{equation} \label{eq9} 
\Delta \Omega =\left(\rho_{l} C_{0} -1\right)\phi_{0} +\frac{1}{2} C_{0} \rho_{l} \phi_{0} ^{2} +\frac{1}{2} 
               \sum_{\vec{G}}\vec{C}_{2} \left(\vec{G}\right)\phi _{\vec{G}} ^{2}   
\end{equation} 

$\Delta \Omega =0$ is the second thermodynamic condition of freezing. Since $\phi _{0} $and $\phi _{\vec{G}} $ are all positive, $\Delta \Omega $ 
can be zero only if $\left(\rho _{l} C_{0} -1\right)\phi _{0} +\frac{1}{2} C_{0} \rho _{l} \phi _{0} ^{2} $ is compensated by 
$\frac{1}{2} \vec{C}_{2} \left(\vec{G}\right)\phi _{\vec{G}} ^{2} $. The above expressions are all for one component liquid. However, as shown by 
Chaudhuri \textit{et al.}~\cite{chandan_prl}, the same expression may be used with an effective $C_{0} $ and effective $\vec{C}_{2} \left(\vec{G}\right)$. 
The order parameters themselves can be determined by a set of non-linear equations, as shown by Rice \textit{et al.}~\cite{xu_rice_pre_08,xu_rice_pre_11}.
Interestingly, in many earlier studies the approach of $\rho _{l} \, \vec{C}_{2} \left(\vec{G}\right)$ to unity was considered as the limit of 
stability of the liquid towards freezing. A different approach was adopted by Rice \textit{et al.}~\cite{bcr_1}. In the present analysis, 
$\rho_{S}\vec{C}_{2}\left(\vec{G}\right)$ approaches unity.
\begin{figure}
\includegraphics[width=0.4\textwidth]{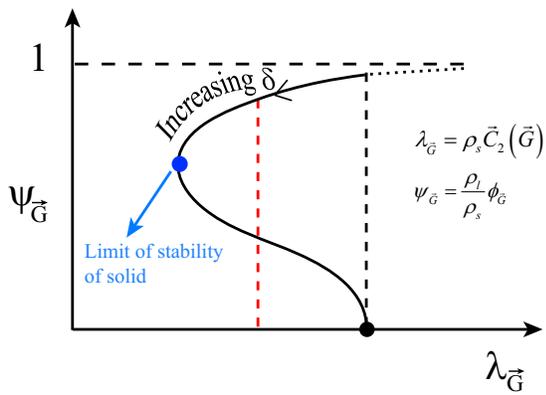}
\caption{\label{fig1}Two order parameter ($\phi_{0}$ and $\phi_{\vec{G}}$) scaled phase diagram plotted in scaled units. The two end points denote two 
limits of stability. The first terminal point gives the limit of stability of the solid, as discussed here, and is connected to terminal solid density.
The second terminal point at $\lambda_{\vec{G}}=1$ gives the limit of stability as discussed earlier by Rice \textit{et al}. and even earlier 
by Lovett~\cite{lovett_jcp_77}.}
\end{figure}
The scaled phase diagram with scaled variables is shown in Fig.~\ref{fig1}. Note this diagram shows a curious character. The scaled phase diagram 
is just like a van der Waals loop with terminal points that may be identified as the two spinodal points of the solid-liquid phase transition. 
The first turn is the one that corresponds to the limit of stability of the solid, the second turn at larger density should correspond to the limit 
of stability of the liquid. Maxwell tie line (given by$\Delta \Omega =0$) is constructed to find coexistence under a given thermodynamic condition.
This enforces the additional thermodynamic condition of equality of pressure between the fluid and the solid branches. Eq.~\ref{eq7} and hence the 
scaled phase diagram have been obtained under the assumption of equality of chemical potential between the liquid and solid phases. Coexistence 
is found by imposing equality of grand thermodynamic potential (see Eq.~\ref{eq9}).

Along X axis, $\lambda _{\vec{G}} $ can correspond to multitudes of $\rho _{S} $ and $\rho _{l} $ values within the region of scaled phase diagram. 
This is because $\lambda _{\vec{G}} $ depends intricately on both $\rho _{S} $ and $\rho _{l} $ values through direct correlation function which is 
evaluated with $\rho _{l} $, but at the reciprocal lattice vector $\overrightarrow{G}$ which depends on the solid density $\rho _{S} .$Equality of 
pressure for grand canonical ensemble picks out unique values of $\rho _{S} $ and $\rho _{l} $. The scaled phase diagram is generated by using same
chemical potential, but the condition of equality of pressure needs to be imposed to obtain the thermodynamic transition parameters.

As the polydispersity parameter $\delta $ increases, freezing shifts continuously to higher volume fraction (and pressure). $C_{0} $ increases 
with density. So, fractional density change decreases to compensate for this as neither $\vec{C}_{2} \left(\vec{G}\right)$ nor $\phi _{\vec{G}} $ 
undergoes such sharp variation. $\phi _{0} $ decreases as $\rho _{s} $ approaches$\rho _{l} $.

There have been several discussions~\cite{bcr_2,bcr_3,bcr_4,bcr_5} that address the limit of freezing. While Lovett~\cite{buff_jcp_80} discusses this 
as a spinodal decomposition, Bagchi-Cerjan-Rice finds this as the last point of the liquid state where the liquid can transform to solid. Beyond
this point no solution to integral equation theory is available. Fig.~\ref{fig1} thus provides two densities of the liquid: one equilibrium transition
density (with $\Delta \Omega =0$) and another terminal density which has the character of a spinodal point. Given the approximate nature of treatment,
it is important to compare its predictions with other estimates.

\section{Simulation details}
We study all simulations using standard Molecular dynamics (MD) and Monte Carlo (MC) techniques supplemented by a particle swap algorithm. Our 
system consists of 500 particles in a periodically repeated cubic box with volume V having Gaussian distribution of particle diameters $\sigma $,
\begin{equation} \label{eq10} 
P\left(\sigma \right)=\frac{1}{\sqrt{2\pi d^{2} } } \exp \left[-\frac{1}{2} \left(\frac{\sigma -\bar{\sigma }}{d} \right)^{2} \right]. 
\end{equation} 

The dimensionless polydispersity index is defined as $\delta =\frac{d}{\bar{\sigma }} $, where $d$ is standard deviation of the distribution and 
$\bar{\sigma }$ is the mean diameter. The particles interact with LJ potential with a constant potential depth. The potential for a pair of 
particles ${i}$ and ${j}$ is cut and shifted to zero, at distance $r_{c} =2.5\sigma _{ij}$, where $\sigma_{ij} =(\sigma _{i} +\sigma _{j} )/2$.
We study for different polydispersity indices starting from $\delta $ = 0 up to $\delta $ = 0.20 with an increment of 0.01. Both MD and MC 
simulations have been employed in NVT, NPT, and NVE ensembles. The MC simulations are aided with particle swapping to enable a faster equilibration 
in the solid phase. In the case of NVT and NVE ensembles, it is customary to monitor the volume fraction ($\varphi $) instead of monitoring 
the density of the system. The standard conjugate gradient method~\cite{numerical_recipe} has been employed to obtain the inherent structures along 
an MD trajectory. We compute the average inherent structure (IS) energy by varying polydispersity. MC simulations with larger system size 
(2048 particles) are carried out to check the finite size effects and the overall physical picture remains the same upon variation of system size.

\section{Results and discussions}
\subsection{ Hansen-Verlet rule of crystallization in freezing of Lennard-Jones polydisperse system}
\begin{figure}[htbp!]
\vspace{15pt}
\includegraphics[width=0.45\textwidth]{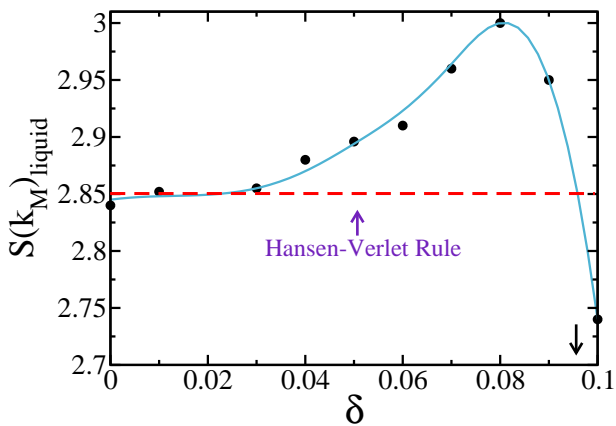}
\caption{\label{fig2}The variation of first peak maximum of the liquid structure factor with different polydispersity indices ${\delta }$. 
The structure factor of the liquid phase for different ${\delta }$ is calculated for the corresponding co-existence volume fraction
${\varphi }$ at temperature $T^*= 1$. The red dotted line signifies the predicted value obtained from Hansen-Verlet rule of crystallization. 
Please note that black arrow shows the structural transformation at polydispersity ${\delta }$ =0.095 predicted by structure factor 
analysis that agrees well with all other analyses (inherent structure, free energy, Lindemann criterion) presented here.}
\end{figure}

In our earlier work~\cite{sarkar_pre} we have obatined the solid-liquid coexistence lines for different polydispersity from free energy calculation 
using umbrella sampling. In the present work, along the coexistence line for a particular co-existence volume fraction $\varphi $, we calculate 
the structure factor of the liquid phase against polydispersity index $\delta $. The first peak maxima of the liquid structure factors 
$S(k_{M})_{liquid}$ for different polydispersity indices are shown in Fig.~\ref{fig2} The plot shows that the first peak 
maximum increases with polydispersity till $\delta $ =0.08 and then it starts decreasing with $\delta $. The red dotted line in the figure 
shows the actual line predicted by the Hansen-Verlet rule of crystallization where the value of the structure factor is exactly 2.85. 
The structural transformation happens at polydispersity $\delta $ =0.095 as shown by black arrow in Fig.~\ref{fig2}. Note that at 
temperature $T^{*} = 1$ and at polydispersity index $\delta $ = 0.10, the value of the structure factor $S(k)$ of the liquid phase 
is 2.74 which is below the value of 2.85 needed for freezing predicted by structure factor analysis. 
\subsection{Lindemann criterion in melting of polydisperse system}
This phenomenological criterion was put forward by Lindemann in 1910. The idea behind the theory was the observation that the average 
amplitude of thermal vibrations increases with increasing temperature~\cite{lindemann}. Melting initiates when the amplitude of vibration becomes 
large enough for adjacent atoms to partly occupy the same space. The Lindemann criterion states that melting is expected when the 
root mean square vibration amplitude exceeds a threshold value. Hence one can write
\begin{equation} \label{eq11} 
L=\frac{\sqrt{\left\langle \mathop{\overrightarrow{r}}\nolimits_{i}^{2} \right\rangle } }{a} , 
\end{equation} 
where $L$ is the Lindemann parameter for the associated polydisperse system and ${a}$ is the mean distance between the 
particles. This threshold value of Lindemann constant is $L\approx 0.1$for melting of crystals~\cite{nelson} of point-like particles in three 
dimensions, but it may vary between 0.05 and 0.20 depending on different factors like nature of inter particle interactions, 
crystal structure, and magnitude of quantum effects.

\begin{figure}[t!]
\includegraphics[width=0.45\textwidth]{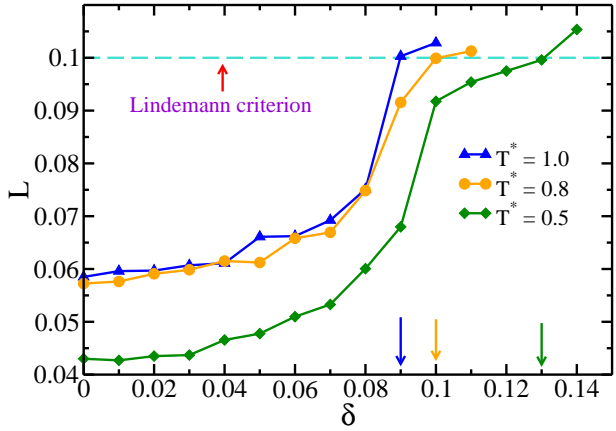} 
\caption{\label{fig3} Variation of Lindemann parameter ($L$) with respect to polydispersity indices (${\delta }$). Please note that 
with increase in polydispersity, the value of $L$ also increases and for temperature $T^{*} =1.0$, the value of $L$ crosses the 
threshold value 0.1 at ${\delta }$ = 0.09 (as indicated by blue arrow), for $T^{*} =0.8$, $L$ crosses the threshold value 
at ${\delta }$ = 0.10 (as indicated by orange arrow) and for $T^{*} =0.5$, $L$ crosses the threshold value at 
${\delta }$ = 0.13 (as indicated by green arrow). Note that blue, orange and green arrows signify the onset of melting 
process at ${\delta }$ = 0.09, 0.10 and 0.13 corresponding to the temperatures $T^*$= 1.0, 0.8 and 0.5 respectively.}
\end{figure}

Fig.~\ref{fig3} shows the variation of Lindemann parameter ($L$) with respect to polydispersity indices ($\delta $) for three different 
temperatures $T^{* }$= 1.0, 0.8 and 0.5 for a constant volume fraction $\Phi $ = 0.58. The plot shows that the value of $L$ increases 
with the increase of $\delta $. For temperatures $T^*$=1.0, 0.8 and 0.5, $L$ crosses the threshold value 0.1 at $\delta $ = 0.09, 0.10 
and 0.13 respectively. This signifies the onset of melting processes at $\delta $ = 0.09, 0.10 and 0.13 for $T^*$=1.0, 0.8 and 0.5 
respectively in the Lennard-Jones polydisperse solid, when the amplitude of the root mean square vibration exceeds a threshold 
value taken as $L$ = 0.1. 

\subsection{Prediction of transition polydispersity from inherent structure analysis}
\subsubsection{Correlation between transition polydispersity and average inherent structure energy}
In polydisperse hard sphere fluids attractive interaction potential is not taken into account. Hard sphere system shows reentrant 
melting of the solid and also above a threshold value of polydispersity (known as terminal polydispersity) hard sphere does not 
form solid phase under any condition. On the other hand, for LJ fluid model system, the attractive potential is taken into account 
and hence it is interesting to study the influence of the attractive part of potential on the terminal polydispersity.

\begin{figure}[t!]
\includegraphics[width=0.45\textwidth]{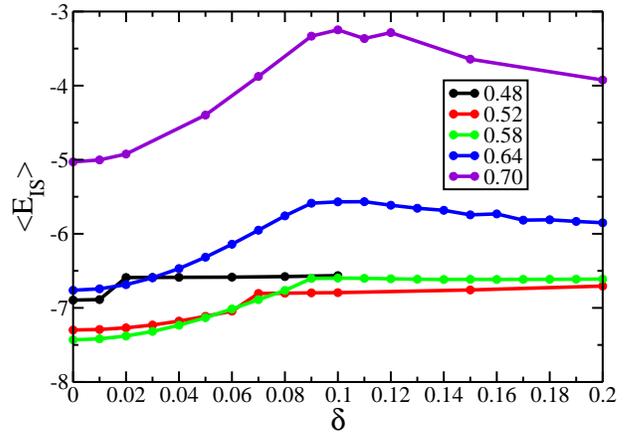}
\caption{\label{fig4} Variation of average IS energy with respect to polydispersity index at constant temperature $T^*$ = 1.0 for different 
volume fractions ${\Phi }$ = 0.48, 0.52, 0.58, 0.64, 0.70. The average IS energy increases with increase in polydispersity 
until transition polydispersity, when the dependence changes sharply. Note that at low volume fraction ${\Phi }$, the transition 
occurs at lower value of ${\delta }$ and at higher ${\Phi }$, solid phase is stable up to higher ${\delta }$ value.}
\end{figure}

As shown in Fig.~\ref{fig4}, the average IS energy is computed with increase in polydispersity for different volume fractions 
$\Phi $ = 0.48, 0.52, 0.58, 0.64, 0.70 at constant temperature $T^* = 1.0$. The average IS energy increases almost quadratically 
with polydispersity up to a polydispersity index at which structural transition takes place. This polydispersity index is known as 
``transition polydispersity,'' beyond which the crystalline solid phase is no longer stable and amorphous phase exists. 
We achieve the transition polydispersity $\left(\delta_{t} \right)$ for different volume fractions using IS analysis. The crossover 
of the two slopes corresponding to each $\Phi $ signifies the transition polydispersity where the structural transition from FCC to 
amorphous state takes place. Table~\ref{table1} shows the transition polydispersities $\left(\delta _{t} \right)$ for the melting 
transition of the Lennard-Jones polydisperse system corresponding to different values of volume fractions and for constant temperature $T* = 1.0$. 

\begin{table}
\caption{\label{table1}Numerical values of transition polydispersity for the melting transition of Lennard-Jones polydisperse systems. 
The first column represents the volume fraction $\Phi$ and second column represents the transition polydispersity ($\delta_{t} $) 
correspond to each $\Phi $.}
\begin{center}
\begin{tabular}{p{0.5in} p{0.5in}} \hline 
${\Phi}$ & $\mathbf \delta_{t}$ \\ \hline
\hline
0.48 & 0.02 \\ %\hline
0.52 & 0.07 \\ %\hline 
0.58 & 0.09 \\ %\hline 
0.64 & 0.10 \\ %\hline 
0.70 & 0.11 \\ \hline 
\end{tabular}
\end{center}
\vspace{-15pt}
\end{table}

\begin{figure*}
\subfigure{\label{fig5a}
\includegraphics[width=0.35\textwidth]{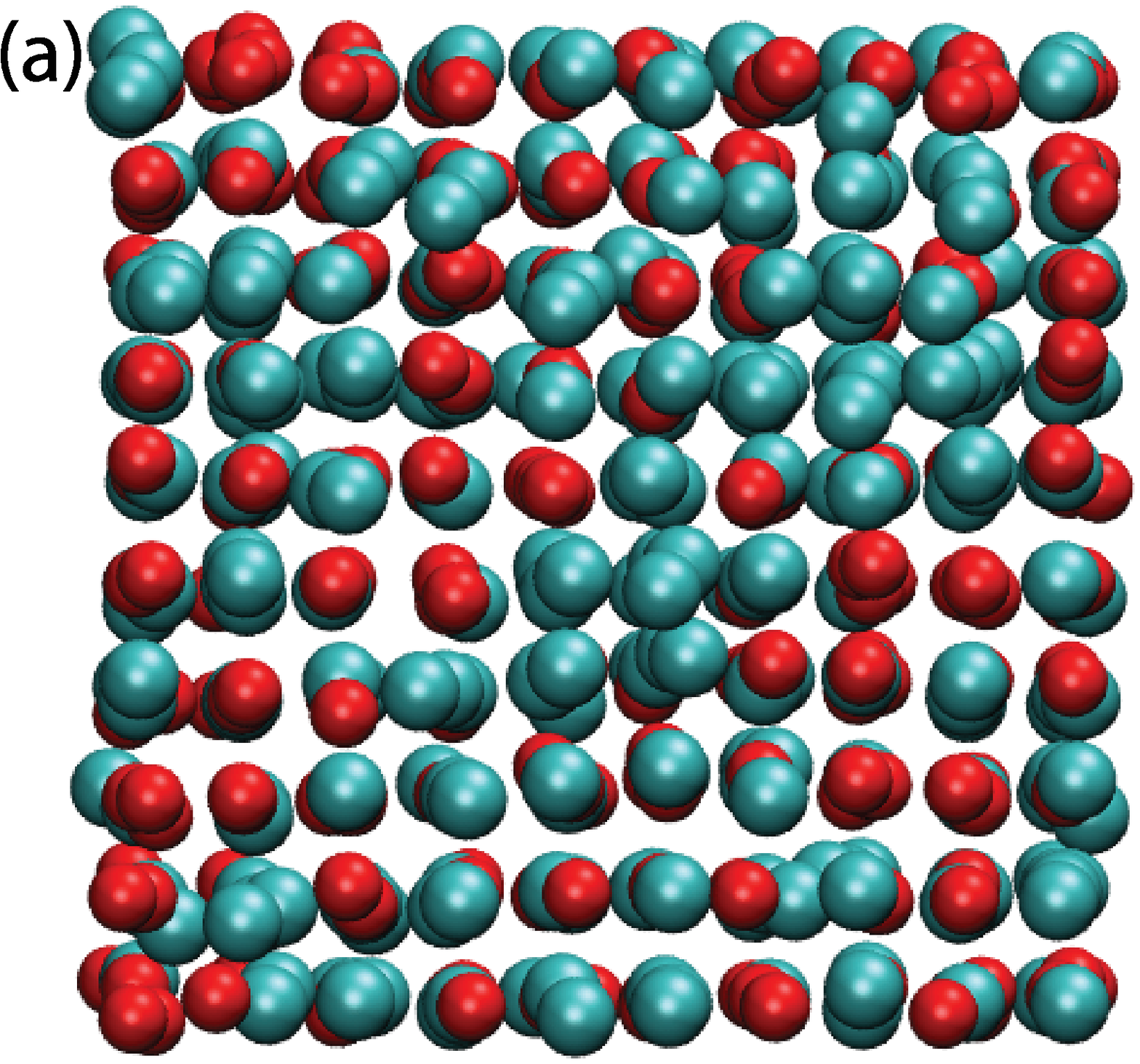}}
\hspace{25pt}
\subfigure{\label{fig5b}
\includegraphics[width=0.35\textwidth]{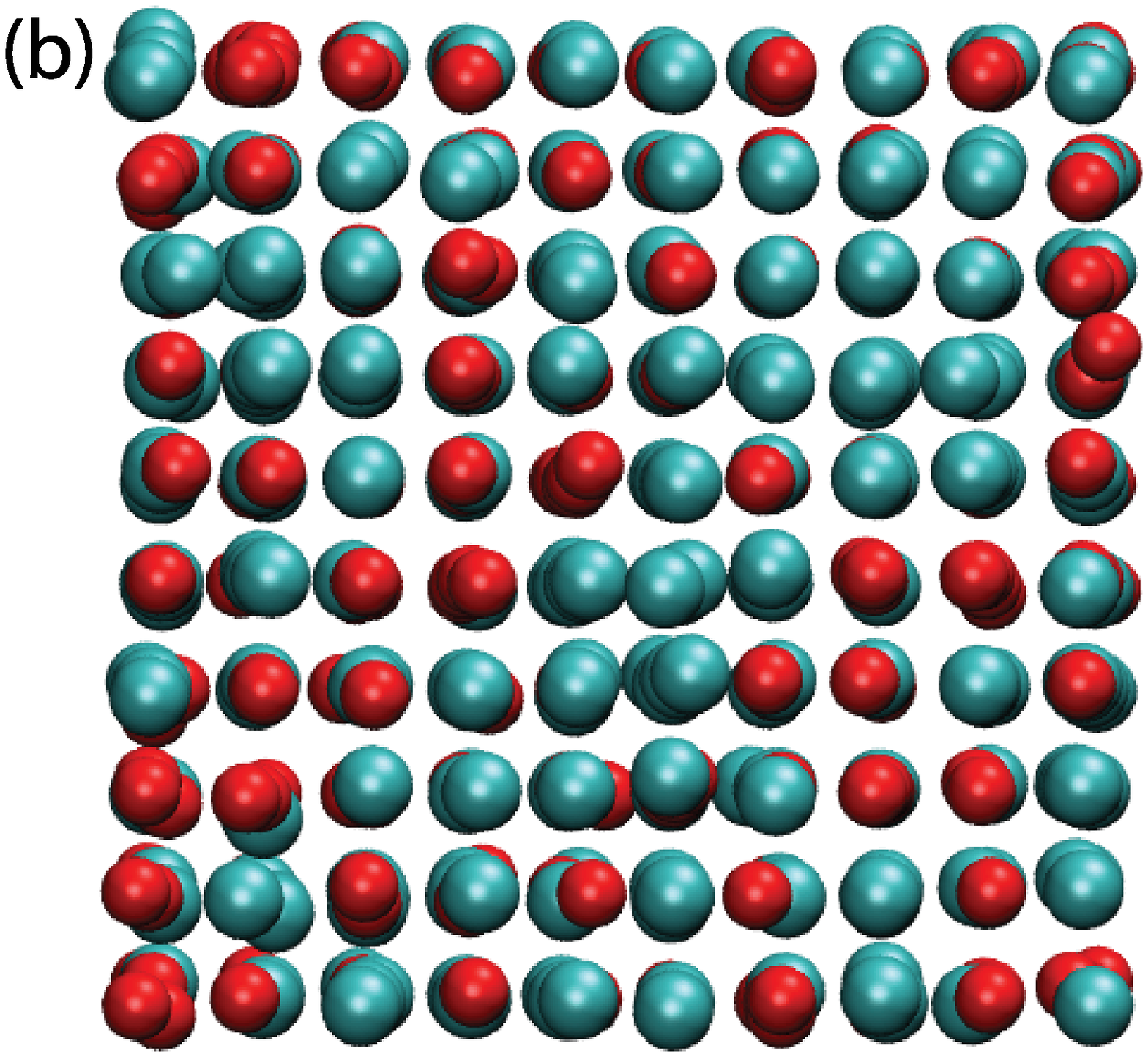}}\\
\subfigure{\label{fig5c}
\includegraphics[width=0.35\textwidth]{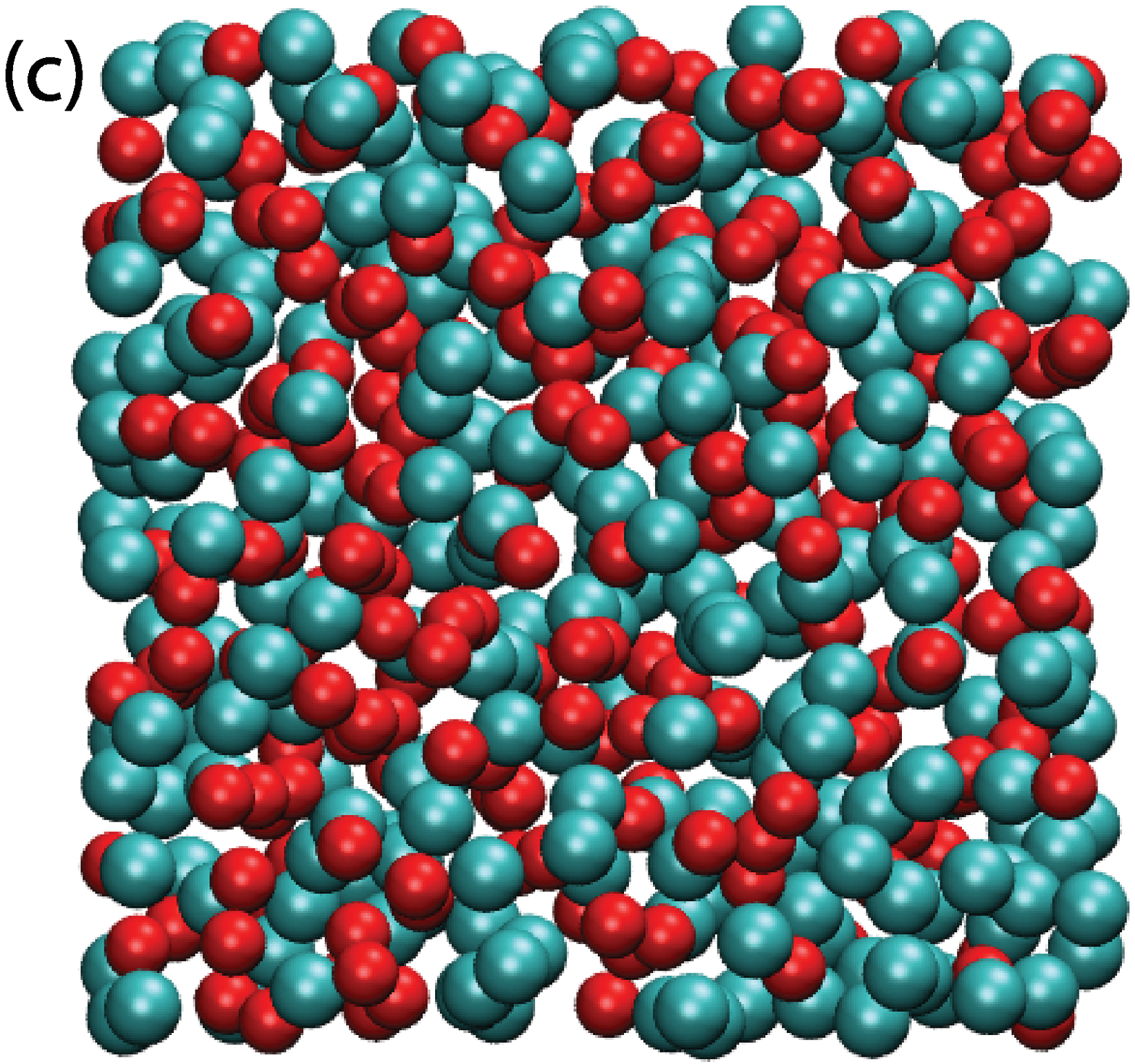}}
\hspace{25pt}
\subfigure{\label{fig5d}
\includegraphics[width=0.35\textwidth]{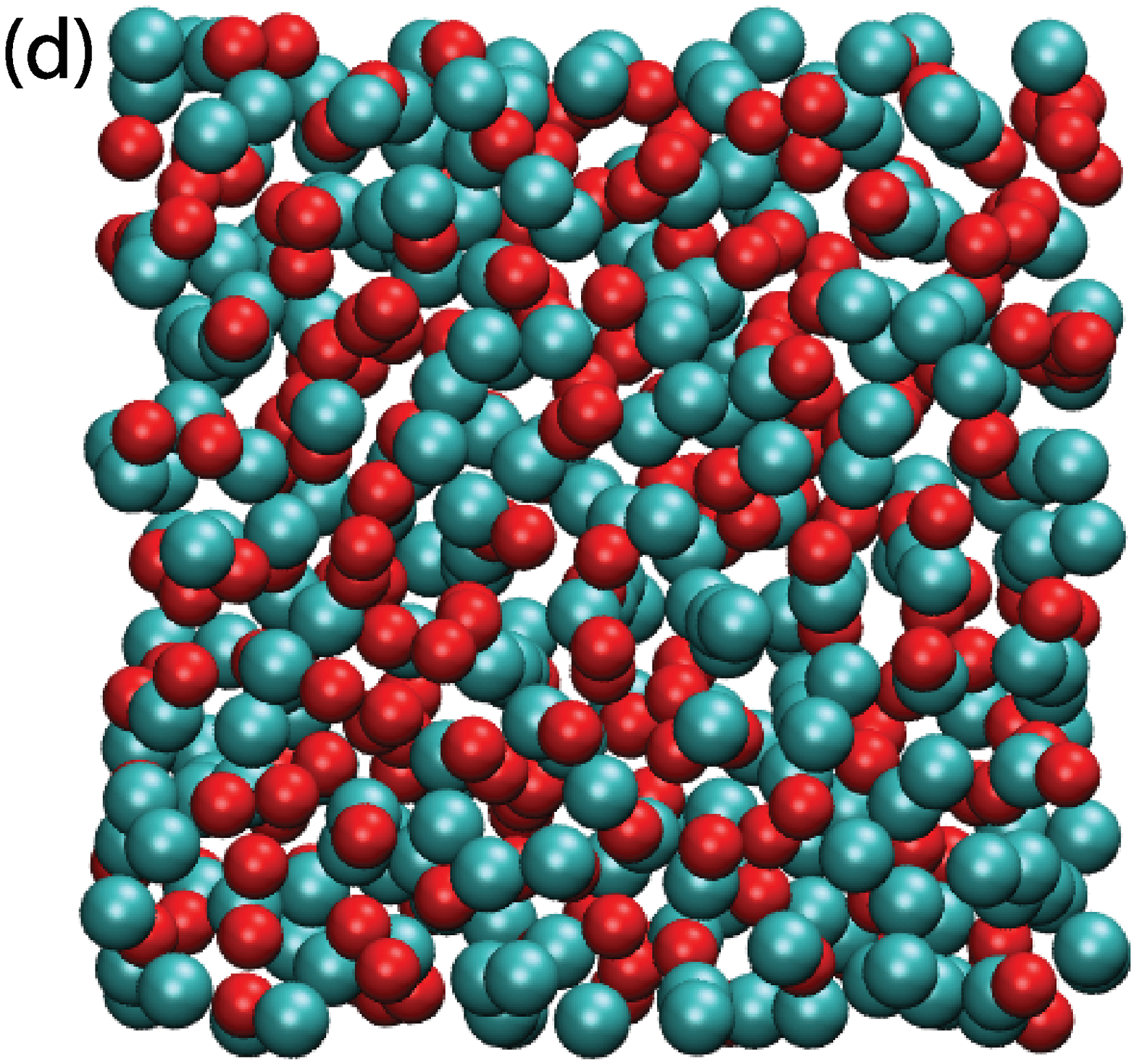}}
\caption{\label{fig5}(Color online) Snapshots showing the spatial positions of small particle in red (dark gray) and big particle in 
cyan (light gray) color in the parent structure along with its corresponding inherent structure at different polydispersity 
indices. Please note that the parent structure is at temperature T*=1 and ${\Phi }$ = 0.58. (a) Parent structure corresponds
to the polydispersity index 0.08. Note the crystal-like structure of the parent polydisperse system (b) Inherent structure at 
${\delta }$ = 0.08 shows the more structured formation than the corresponding parent structure (c) Snapshot shows parent 
structure at ${\delta }$ = 0.09 and (d) corresponding inherent structure at ${\delta }$ = 0.09. Note that fig. 5(c, d) 
depict amorphous/glass like formation in the parent and inherent structure at ${\delta }$ = 0.09.}
\end{figure*}
Difference in the polydispersity index ($\delta $) dependence of the inherent structure (IS) energy in the crystalline solid and in 
the amorphous phase is understandable in terms of the theory of elasticity in solids~\cite{landau}. In the presence of a strain field, the free 
energy can be written as,
\begin{equation} \label{eq12} 
F=F_{0} +\frac{1}{2} \lambda u^{2} _{ii} +\mu u_{ik} ^{2} , 
\end{equation} 
where $\lambda $ and $\mu $ are called \textit{Lam\'{e} coefficients}.
\begin{equation} \label{eq13} 
K=\lambda +\frac{2}{3} \mu . 
\end{equation} 
Here $K$ and $\mu $are bulk modulus and modulus of rigidity respectively.

Now in the crystalline solid, polydispersity introduces strain field $u_{ik} $which is randomly distributed in the solid and strongly 
dependent on $\delta $. As $\delta $ increases, strain field also increases. The quantitative estimate in the free energy can be 
achieved from the density-functional theory of elasticity~\cite{marko_prb,lipkin_jcp_85}. On the other hand, the particles in the amorphous 
phase can undergo rearrangements to reduce the strain which explains the weak dependence of the IS energy with $\delta $ in the amorphous phase.
This allows us to estimate quantitatively the criterion of transition from IS analysis. Another point to notice is that the change of the IS energy at 
the cross over depends both on temperature and volume fraction. As the volume fraction is lowered, the effect of the strain field on the IS 
energy of the solid decreases. This describes the reason of different transition polydispersity indices $\left(\delta _{t} \right)$for 
different volume fractions. 

\subsubsection{Structural patterns of parent structure and corresponding inherent structure at different polydispersity indices}
The transition polydispersity for the melting transition of the Lennard-Jones polydisperse system needs further exploration. 
We study the structural aspects at different polydispersity indices. We categorize the particles into two sub ensembles; particles
with diameter less than 1.0 are termed as small particle and rest of the particles as big particle. In Fig.~\ref{fig5} we represent the 
snapshots showing the spatial positions of small particle in red and big particle in cyan color in the parent structure along 
with its corresponding inherent structure at different polydispersity indices. The snapshots of the parent liquid have been taken 
at reduced temperature $T^*=1.0$, and volume fraction $\Phi $ = 0.58. Fig.~\ref{fig5a} shows an instantaneous parent structure at 
$\delta $ = 0.08 and Fig.~\ref{fig5b} shows its corresponding inherent structure while Figs.~\ref{fig5c},~\ref{fig5d} depict 
molecular arrangements for the parent structure at $\delta $ = 0.09 and its inherent structure respectively. Fig.~\ref{fig5a} and ~\ref{fig5b}
reveal the crystal-like arrangement of the particles in both parent and inherent structure respectively. Fig.~\ref{fig5c} and~\ref{fig5d}
show instantaneous parent structure and corresponding inherent structure at $\delta $ = 0.09 which seem to be almost homogeneous. These four snapshots 
show structural transition of the present L-J polydisperse system when the system enters from $\delta $ = 0.08 to $\delta $ = 0.09.
\subsection{Relationship among scaled order parameter ($\lambda_{\vec{G}}$) in the scaled phase diagram and coexistence densities of solid and liquid}

\begin{figure}
\includegraphics[width=0.5\textwidth]{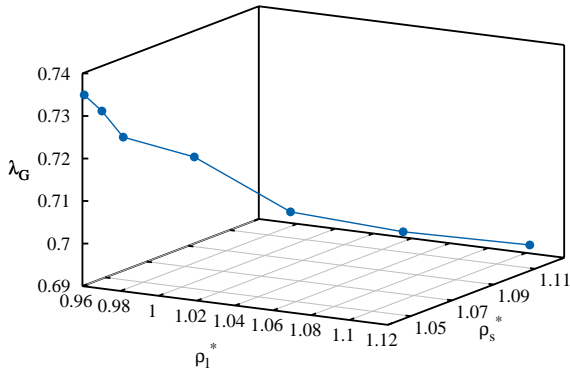}
\caption{\label{fig6} Variation of scaled order parameter ($\lambda_{\vec{G}}$) in the phase diagram with coexistence density
of solid $(\rho_{s}^{*})$ and liquid $(\rho_{l}^{*})$. Note that the value of the scaled order parameter
($\lambda_{\vec{G}}$) decreases as both the coexistence density of solid $(\rho_{s})$ and that of liquid $(\rho_{l})$ increase.}
\end{figure}

With the increase in polydispersity indices, the coexistence density of the solid $(\rho _{s}^{*} )$approaches that of the liquid 
$(\rho _{l}^{*} )$ and the value of the direct correlation function at its first peak decreases. Hence, the value of the order 
parameter ($\lambda _{\vec{G}} $) in the scaled phase diagram decreases with the increase in coexistence density of solid $(\rho _{s}^{*} )$ 
and that of liquid $(\rho _{l}^{*} )$ as shown in Fig.~\ref{fig6}.
That is, in the phase diagram given by Fig.~\ref{fig1}, one moves toward left along the upper branch. Since no transition is possible beyond 
the spinodal point, this point can be identified with the instability point of Lindemann melting criteria.
\subsection{Discussion}
In our present study we are interested in melting of the solid due to increase in polydispersity, as elaborated by the IS analysis.
As the polydispersity increases, several things occur that determine the position of the coexistence. First, the density of the solid 
at coexistence approaches that of the liquid. Also, the crystalline order of the solid decreases progressively which is reflected in 
the lower value of the order parameters. Thus, the value of the ordinate (that gives scaled order parameter,$\psi _{\vec{G}} $) in 
the scaled phase diagram (the upper solid line) decreases as polydispersity increases. At the same time, value of the abscissa also 
decreases because the density of the solid approaches that of the liquid and the value of the direct correlation function at its first 
peak decreases. All these lead the system towards the first terminal point which can thus be identified as the \textit{limit of stability
of the polydisperse solid with respect to the liquid}. The results obtained for the melting transition parameters of the Lennard-Jones 
polydisperse solids are shown in Table~\ref{table2}. Interestingly, the value of the polydispersity index (for a constant volume fraction) 
at which melting is predicted agrees well with Lindemann criteria of atom displacement.

In order to explore the mechanical stability of a metastable solid through inherent structure analysis, the potential energy of the solid 
must be a minimum. We find that beyond a given polydispersity index (at a constant volume fraction) no such minimum, indicating that 
amorphous state represents the true ground state of the system.

From the liquid side, we can get an estimate of the transfer of stability from liquid to solid state by (a) Ramakrishnan and Yussouff theory
~\cite{tvr_prb_79} (b) empirical Hansen-Verlet rule~\cite{hansen_pr}. Both again give values of liquid and solid density those are 
in agreement with inherent structure analysis and integral equation based theoretical analysis of liquid-solid transition.

\begin{table}
\caption{\label{table2} Comparison of values of melting parameters. The first column represents value of polydispersity index (${\delta }$) 
and second column represents corresponding critical pressure $(P^{*})$ at coexistence of solid and liquid phases. Third and fourth columns 
represent critical density of solid $(\rho_{s}^{*})$ and its volume fraction $(\phi_{S})$ respectively. Fifth and sixth columns represent 
critical density of liquid $(\rho_{l}^{*})$ and its volume fraction $(\phi_{l})$ respectively. The value of $(\phi_{0})$ represents 
fractional density change on melting transition. The value of $S\left(k_{M}\right)_{L}$ shows the value of the liquid structure factor 
at the first peak. In the last column $\lambda_{\vec{G}}$ represents the value of the scaled order parameter.}
\begin{center}
\begin{tabular}{l|l|l|l|l|l|l|l|l} \hline
\textbf{${\mathbf \delta }$} & \textbf{$P^{*} $ } & \textbf{$\rho _{s}^{*} $ } & \textbf{$\phi _{S} $ } & 
\textbf{$\rho _{l}^{*} $} & \textbf{$\phi _{l} $ } & \textbf{$\phi _{0} $} & \textbf{$S\left(k_{M} \right)_{L} $} & 
\textbf{$\lambda _{\vec{G}} $} \\ \hline \hline
\textbf{0.0} & 6.58 & 1.039  & 0.544 & 0.961 & 0.503 & 0.075 & 2.839 & 0.735 \\ \hline 
\textbf{0.01} & 6.73 & 1.042 & 0.546 & 0.967 & 0.507 & 0.072 & 2.852 & 0.731 \\ \hline 
\textbf{0.03} & 7.10 & 1.044 & 0.547 & 0.976 & 0.511 & 0.065 & 2.855 & 0.725 \\ \hline 
\textbf{0.04} & 8.28 & 1.058 & 0.554 & 0.998 & 0.523 & 0.058 & 2.879 & 0.719 \\ \hline 
\textbf{0.06} & 10.40 & 1.074 & 0.563 & 1.031 & 0.540 & 0.040 & 2.911 & 0.705 \\ \hline 
\textbf{0.07} & 12.97 & 1.097 & 0.575 & 1.065 & 0.558 & 0.029 & 2.961 & 0.698 \\ \hline 
\textbf{0.08} & 16.30 & 1.124 & 0.588 & 1.102 & 0.577 & 0.019 & 3.001 & 0.692 \\ \hline 
\textbf{0.09} & - & - &  & - & - &  & 2.949 & - \\ \hline 
\textbf{0.10} & - & - &  & - & - &  & 2.740 & - \\ \hline 
\end{tabular}
\end{center}
\vspace{-10pt}
\end{table}
We note that if we apply Ramakrishnan-Yussouff theory to find the freezing density with an effective direct correlation function, 
then we obtain values that are close to the values given in Table~\ref{table2}. As evident from Table~\ref{table2} (also indicated 
in Fig.~\ref{fig1}), with increasing polydispersity index $\delta $, density of liquid at solid- liquid transition increases. This leads 
to an increase in the first peak of the liquid structure factor, $S(k)_{Liquid}$. At the same time, value of the scaled order 
parameter ($\lambda_{\vec{G}} $) decreases because the density of the co-existence solid increases. This is 
the direct consequence of the sharp fall in the fractional density change ($\phi_{0}$) on melting-freezing. As $\delta $ 
increases, both of the changes in $\lambda_{\vec{G}}$ and $\psi_{\vec{G}}$ values drive the system to the turning 
point/spinodal point (as shown in Fig.~\ref{fig1}). Below the particular value of $\lambda_{\vec{G}}$ at spinodal point, no 
liquid-solid transition is possible.
\section{Conclusion}
\vspace{-5pt}
The terminal polydispersity has not been studied earlier in the Lennard-Jones polydisperse system using non-linear integral equation 
theories of freezing. We discuss how certain features of the scaled phase diagram may be used to explain the terminal polydispersity 
in polydisperse system. This analysis implies that at high polydispersity, the equality of thermodynamic potential between solid and 
liquid phases can no longer be established. The liquid state has lower chemical potential and the solid phase is not possible beyond 
certain polydispersity. This is manifested in the absence of any solution of the non-linear integral equations. The same feature is 
reflected in the inherent structure analysis - no crystalline solid is obtained in the inherent structure beyond the terminal 
polydispersity. The amorphous state is the global minimum in the potential energy surface.
\begin{acknowledgements}
We would like to thank Prof. Stuart A. Rice and Dr. Rakesh S Singh for discussions. This work was supported 
in parts by grants from DST and CSIR (India). B.B. thanks DST for support through J.C. Bose Fellowship. 
\end{acknowledgements}
%\bibliography{manuscript_pds}
%
\end{document}